\documentstyle[prb,twocolumn,epsfig,aps]{revtex}

\begin{document}

\draft

\title{Theory for the coupling between longitudinal phonons and intrinsic       
Josephson oscillations in layered superconductors}

\author{Ch. Helm, Ch. Preis, Ch. Walter, and J. Keller}
\address{Institut f\"ur Theoretische Physik, Universit\"at Regensburg, \\  
D-93040 Regensburg, Germany \\
(submitted to Phys.Rev.B)} 


\maketitle

\begin{abstract}

In this publication a microscopic theory for the coupling of 
intrinsic Josephson oscillations in layered
superconductors with  longitudinal c-axis-phonons is
developed. It is shown that the influence of lattice vibrations on the
$c$-axis transport can be fully described by introducing an  effective
longitudinal dielectric function $\epsilon^L_{\rm ph} (\omega)$. Resonances 
in the $I$-$V$-cha\-rac\-ter\-is\-tic appear   at van Hove singularities of 
both acoustical and optical longitudinal phonon branches. This provides a natural
explanation of the  recently discovered subgap structures in the
$I$-$V$-characteristic of highly anisotropic cuprate superconductors.  
The effect of the phonon dispersion on the damping of these resonances and 
the coupling of Josephson oscillations in different resistive
junctions due to phonons are discussed in detail.

\end{abstract}


\narrowtext

\section{Introduction}

The $c$-axis transport in the highly anisotropic
cuprate superconductors Tl$_2$Ba$_2$Ca$_2$Cu$_3$O$_{10+\delta}$ (TBCCO) and 
Bi$_2$Sr$_2$CaCu$_2$O$_{8+\delta}$ (BSCCO) can well be described by a stack of
Josephson junctions between the superconducting CuO$_2$-mul\-ti-lay\-ers.
This intrinsic Josephson effect can be seen in the multi-branch structure of
the $I$-$V$ characteristic,  each
branch corresponding to a well-defined number of junctions in the
resistive  state.\cite{kleiner1,kleiner2} The intrinsic Josephson
effect is observed also in the behaviour of the material in external magnetic 
fields and under microwave irradiation.\cite{kleiner1,kleiner2}

Recently subgap structures in the  $I$-$V$-cha\-rac\-te\-ris\-tic have been
discovered as intrinsic properties of the 
material,\cite{schlengal,yurgenss,seidel}
which could be explained 
by the coupling between the intrinsic Josephson oscillations and 
phonons.\cite{wir1,wir2,wir3}
This interaction is mediated by the oscillating electric field  
in the Josephson junction, which excites vibrations of charged ions in the
material.  
In our previous investigations \cite{wir1,wir2} we used a
simple model of a system of damped harmonic oscillators in order to describe 
the dynamics of oscillating
ions in  the barrier. We were able to derive an analytic expression for the
dc-current density $j(V)$ as function of the dc-voltage $V$ for one resistive
junction 
\begin{equation}
j(V)= j_{qp}(V) + \frac {j_c}{2} \frac{\omega_p^2}{\omega^2}
\frac{\epsilon_2 + \frac{\sigma}{\omega \epsilon_0}}{\epsilon_1^2 +
(\epsilon_2 + \frac{\sigma}{\omega\epsilon_0})^2}. \label{IV-curve}
\end{equation}
where the voltage $V$ is related to the Josephson oscillation frequency $\omega$ by
$V=\hbar \omega/(2e)$ and $\epsilon(\omega)=\epsilon_1(\omega) + i
\epsilon_2(\omega)$ is the dielectric 
function of the oscillating ions. From
this result it can be seen that the $I$-$V$-curve has a maximum at the
frequency (voltage) where the real part $\epsilon_1$ of the phonon-dielectric
function vanishes, which corresponds to a longitudinal eigenfrequency of the
phonon system. 

With appropriate values for the Josephson plasma
freqency $\omega_p$, the quasiparticle conductivity $\sigma$, the critical
current denstiy $j_c$, and the frequencies, dampings  and oscillator strengths 
of phonons in the dielectric function we were able to fit the experimental
results for the sub-gap structures in the $I$-$V$-curves perfectly.  In
addition to this, the identification of the maxima of the structure  with
phonon frequencies provides a natural explanation, why the position of  the
resonances is completely independent of temperature, magnetic field and  the
geometry of the probe. The voltage positions of these resonances  can  also be
related to structures in optical experiments,\cite{Zett,Buck} in particular 
to the reflectivity for oblique incidence\cite{tsvetkov}.

Despite the clear evidence of the effect a general theory for the coupling
between Josephson oscillations and phonons 
is still missing and shall be presented in the following. In particular, in
Refs.\ \onlinecite{wir1,wir2} the phonon frequencies  and
the oscillator strengths in the dielectric function were not specified.  
The clarification of this point and the extension to a
general lattice dynamical model including acoustical and optical branches
will be one of the main topics of the following
investigation.

Another topic concerns the coupling of Josephson oscillations in different
resistive junctions. We will show how phonons lead to a coupling of the
phases of Josephson oscillations in different resistive junctions. 
This will be discussed in detail for two resistive junctions and general
results will be given for large stacks of resistive junctions. 
A phase-locking in a stack of Josephson junctions is important for applications
of such systems for high-frequency mixers and detectors.  

The excitation of phonons by Josephson oscillations in conventional 
single Josephson junctions has been observed already a long time 
ago.\cite{kinder}
Also in the $I$-$V$
curves of break-junctions of cuprate-superconductors \cite{Pon} structures
due to phonons might have been 
identified. The physical mechanism described here can
also be applied to these cases but our formalism is particularly suited to
treat stacks of Josephson junctions with phonons in the frequency range
between the Josephson plasma frequency and the gap-frequency.  

It is not our intention to explain  the  details of the
$I$-$V$-characteristic of TBCCO and BSCCO by a realistic lattice 
dynamical calculation. This is impossible at the moment for the complicated anisotropic
superconductors with variable doping showing this effect. 
Furthermore this would require a detailed theory of superconductivity 
and the Josephson effect in two- and three-layer systems including information 
about the superconducting bands, charge distribution, and charge
susceptibility  between the layers and inside the
CuO$_2$ planes which is not available yet. Therefore we discuss a simple model
system with superconducting mono-layers where the conduction electron charge
is distributed homogeneously along the layers, and a lattice dynamical model with 
only two phonon bands  showing already the basic features of a full theory
which will be expected also for realistic systems. We start the discussion
with a definition of our model and a short  derivation of the basic Josephson
equations for a stack of Josephson junctions.

\section{Josephson equations for a stack of junctions}

We consider a system of $N$ superconducting layers separated by
insulating barrier material of thickness $d$ forming a stack of Josephson
junctions. We treat the superconducting layers as homogeneous metal sheets
with a uniform electron distribution along the layers. In this paper we treat
only the case of a uniform  tunneling current  with a
constant bias-current and  neglect  magnetic field effects due
to the current flow. In this case all quantities are constant along the
layers. Such an approximation is reasonable for a stack of junctions which
is short with respect to the  Josephson penetration length  but long enough to
neglect finite size effects in the ionic polarisation.

The tunneling current density $j_n$ from layer $n$ to $n+1$ creates 
(two-dimensional) charge density  fluctuations $\delta \rho_n$ on the layers
related by the continuity equation 
\begin{equation}
j_n(t)- j_{n-1}(t) = - \delta \dot \rho_n(t). \label{cont}
\end{equation}
These charge fluctuations create electric fields $E^\rho_n(t)$ (in the
c-direction) in the barrier between layer $n$ and $n+1$  which are constant 
inside 
each barrier and
are related to the charge fluctuations by the Maxwell equation:                
\begin{equation}
\delta \rho_n(t) = \epsilon_0 \Bigl(E^\rho_n(t) - E^\rho_{n-1}(t)\Bigr)
\label{Maxwell1} \end{equation}
or
\begin{equation}
E^\rho_n(t) = \frac{1}{2\epsilon_0} \Bigl( \sum
\limits_{n'\le n} \delta\rho_{n'}(t) -  \sum\limits_{n'>n}\delta\rho_{n'}(t) 
\Bigr). \label{Maxwell2}
\end{equation}

Assuming that the current density $j_n$ for the first and last barrier is 
fixed by the
bias current density $j$, then with help of Eqs.\ (\ref{cont},\ref{Maxwell1}) 
the
tunneling  current density $j_n$ in all the other junctions is                
related to the bias current density $j$ by 
\begin{equation}
j=j_n(t) + \epsilon_0 \dot E_n^\rho(t).
\end{equation}
The last term is the displacement current density related to the charge
fluctuations on the layers. In the following we denote this term by  $\dot
D_n(t) := \epsilon_0 \dot E^\rho_n(t)$ in order to relate the
present results to  the usual notation of the resistively shunted Josephson
junction (RSJ) model, however, one should keep in mind the microscopic origin
of this term.

In the following we approximate the tunneling current  by a
superposition of a Josephson supercurrent density and a 
quasiparticle current density. Then we have for each junction:
\begin{equation}  
j = j_c\sin\gamma_n(t) + j_{\rm qp}(E_n(t)) + \dot D_n(t). \label{RSJ0}
\end{equation} 
The Josephson current density $j_c\sin \gamma_n(t)$ depends on the gauge
invariant phase difference $\gamma_n(t)$
between layers $n$ and $n+1$ at positions $z_n$ and $z_{n+1}$. It is
related to the average total electric field  in the barrier 
\begin{equation}
E_n(t):= \frac{1}{d}\int_{z_n}^{z_{n+1}} E_z(z,t) dz \label{totalE}
\end{equation} 
by the second Josephson equation 
\begin{equation}
\frac{\hbar}{2ed} \dot \gamma_n(t) =  E_n(t) .\label{2Jos}
\end{equation}
Here  small corrections to Eq.\ (\ref{2Jos}) for layered
superconductors which are discussed in Refs.\ \onlinecite{Koy96,wir4} are
neglected. 
For the quasi-particle current density we will use in the following 
an ohmic
form $j_{\rm qp}= \sigma E_n$ with a constant conductivity $\sigma$. The
generalisation to more realistic forms \cite{wir2} is straightforward. 

The crucial point where the phonons come into play is the relation bet\-ween 
the field $D_n=\epsilon_0 E^\rho_n$, which is created by the charge
fluctuations on the superconducting layers alone, and the average electric
field Eq.\ (\ref{totalE}) $E_n= E^\rho_n+ E^{\rm ion}_n$ which contains also 
the averaged field $E^{\rm ion}_n$ created by the ionic displacements in the   
barrier.  This will be discussed in detail below.                                       

Before we do this let us summarize the most important parameters
which characterize the Josephson system: The first one is the (bare) 
Josephson plasma frequency
$\omega_p$ defined by 
\begin{equation}
\omega_p^2 := \frac{2edj_c}{\hbar \epsilon_0}  \; .
\end{equation} 
The second one is the so-called
characteristic frequency defined by
\begin{equation}
\omega_c:= \frac{2eV_c}{\hbar}  \; .
\end{equation}
Here  $V_c$ is the
voltage  where the quasiparticle current density equals the value $j_c$. It
is of the order of the superconducting energy gap and is a measure of the
dissipative properties of the junction. In our simple model with a constant
conductivity we have  $\omega_c= 2edj_c/(\hbar \sigma)$.  The ratio $\beta_c=
\omega_c^2/\omega_p^2$ is the McCumber parameter, which is  $\beta_c\gg 1$
 for the strongly anisotropic cuprate-superconductors. 
Moreover, for these materials there
exist phonons with  frequencies in the range $\omega_p <
\omega_{\rm phon} \ll \omega_c$. 

Typically the time-dependence of the phase difference $\gamma(t)$
can be written in the so-called resistive state  as 
\begin{equation}
\gamma(t) = \theta + \omega t + \delta \gamma(t)
\end{equation}
where $\omega = \langle\dot\gamma\rangle $ is the time-average of the phase
velocity which is non-zero for a junction in the resistive state. It
determines  the dc-voltage $V = \langle E\rangle d = \hbar \langle \dot
\gamma\rangle /(2e)$ across the junction.  In the 
asymptotically stable state and for large
values of the McCumber parameter $\beta_c$ the oscillating part  $\delta
\gamma(t)$ is small and oscillates with the frequency $\omega$. 

\section{Excitation of phonons by Josephson oscillations} 

Now let us turn to the discussion of lattice vibrations.
Quite generally the lattice displacement of an ion of type $\kappa$ with
mass $M_\kappa$, charge $Z_\kappa$ in unit cell $l$ is determined by the
following equation of motion  \begin{equation}
M_\kappa \ddot u_\alpha({\textstyle{l\atop \kappa}}\vert t) + \sum_{l'\kappa'}
\Phi_{\alpha\beta}({\textstyle{l\atop\kappa}{l'\atop \kappa'}}) u_\beta
({\textstyle{l'\atop\kappa'}}\vert t) = Z_\kappa E^\rho_\alpha
({\textstyle{l\atop\kappa}}\vert t). 
\end{equation}
Here $E^\rho_\alpha ({l\atop\kappa}\vert t)$ is the local driving field at the
equilibrium position $\vec R({l\atop\kappa})= \vec R(l) + \vec R(\kappa)$
of the ion generated by the charge fluctuations $\delta \rho_n(t)$ on the
superconducting layers. 
Note that the vibrating ions may be both in
the barrier material and on the superconducting layers. The superconducting
electrons are assumed to move together with the ions  of the layers.

While the RSJ-equations for the phases are highly 
non-linear, the relations between lattice displacements and electric fields
are linear, consequently we may analyse the response for each frequency
$\omega$ separately. 
With a harmonic ansatz of the form 
$u_\alpha ({\textstyle{l\atop\kappa}}\vert t) =
u_\alpha ({\textstyle{l\atop\kappa}}) e^{-i\omega t} $ 
we obtain for the amplitude 
\begin{eqnarray}\label{displacement}
u_\alpha({\textstyle{l\atop\kappa}}) &=& \frac{1}{N} \sum_{\vec q\lambda} 
\sum_{l'\kappa'\beta} 
\frac {e_\alpha (\kappa\vert \vec q\lambda) e_\beta^*(\kappa'\vert\vec
q\lambda)}         {\omega^2(\vec q\lambda)-\omega^2} \\
\nonumber
&\times&\frac{
e^{i\vec q(\vec R({\scriptstyle{l}})- \vec
R({\scriptstyle{l'}}))}}{\sqrt{M_\kappa M_{\kappa'}}}
eZ_{\kappa'}E^\rho_\beta({\textstyle{l'\atop\kappa'}}). 
\end{eqnarray}
Here $\omega^2(\vec q \lambda)$ and $\vec e(\kappa\vert \vec q \lambda)$
are the eigenvalues and eigenvectors of the dynamical matrix
\begin{equation}
\sum_{\kappa'\beta} D_{\alpha\beta}({\textstyle{\vec q\atop\kappa \kappa'}})
e_\beta(\kappa'\vert \vec q \lambda) =
\omega^2(\vec q \lambda)e_\alpha(\kappa\vert \vec q \lambda)
\end{equation}
defined by
\begin{equation}
D_{\alpha\beta}({\textstyle{\vec q \atop \kappa \kappa'}})=
\sum_{l'}\frac{1}{\sqrt{M_\kappa M_{\kappa'}}}
\Phi_{\alpha\beta}({\textstyle{l\atop\kappa}{l'\atop\kappa'}})  
e^{i\vec q(\vec R({\scriptstyle{l'}}) - \vec
R({\scriptstyle{l}}))}.  
\end{equation}
The force-constant matrix contains the
quantum-mechanical  short-range interactions but also the short-range and
long-range Coulomb  interactions (the latter being of the form $\sim q_\alpha 
q_\beta/q^2$) between the ions, but not the fields set-up by the conduction
electrons on the layers.   The eigenfrequencies $\omega(\vec q\lambda)$ are
therefore by construction the {\it bare}  phonon frequencies in the absence of
the conduction electron charge fluctuations $\delta \rho_n(t)$. They include
possible renormalisation by intra-atomic electronic polarisation. 

\begin{figure}
\epsfig{figure=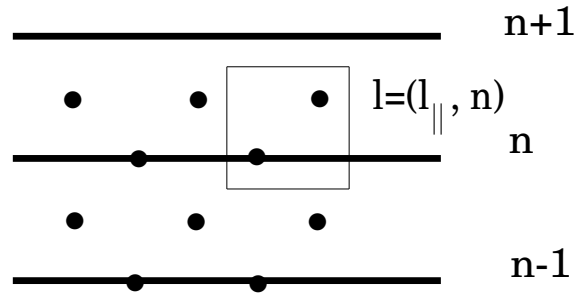, height=4cm}
\caption{Labeling of unit cells. \label{fig1}}  
\end{figure}

For the further discussion it is convenient to label the lattice dynamical unit 
cells by $l=(l_x,l_y,l_z)$ with $l_z=n$ denoting the superconducting layer in
which the lattice cell is contained (see Fig.\ \ref{fig1}). Then the 
z-component of the
position vector $\vec R({\textstyle{l\atop\kappa}})$ for $l=(l_x,l_y,n)$ does
not depend on $l_\parallel=(l_x,l_y)$ and we may write $R_z({l\atop\kappa})=
R_z({n\atop\kappa})= R_z(n) + R_z(\kappa)$. Furthermore  the
origin of the unit-cell may be chosen on the superconducting layer, so that
$R_z(n)=z_n$.  

In our model for the superconductivity in the layers we have to assume that 
the mobile electronic charge is spread out uniformly along the layers, 
because a microscopic theory connecting the superconducting bands and
the atomic
sites in the CuO$_2$-layers is still missing.  
Therefore the field $E_\beta ^\rho({l\atop \kappa})$ does
not depend on the position $x$ along the layer and has only a z-component.
In Eq.\ (\ref{displacement}) only modes with $q_\parallel=0$
appear and we may write for the displacement amplitude in z-direction 
of an ion of type $\kappa$ in barrier $n$:
\begin{eqnarray}  
u_z({\textstyle{n\atop\kappa}})
&=& \frac{1}{N_z} \sum_{q_z\lambda} \sum_{n'\kappa'}  \frac {e_z(\kappa\vert
q_z\lambda) e_z^*(\kappa'\vert q_z\lambda)}
{\omega^2(q_z\lambda)-\omega^2}\\
\nonumber
&\times&\frac{ e^{iq_z(z_{{n}}- 
z_{{n'}})}}{\sqrt{M_\kappa M_{\kappa'}}}
e Z_{\kappa'}E^\rho_z({\textstyle{n'\atop\kappa'}}).  
\end{eqnarray} 

Now let us  specify the  driving field $E^\rho_z({n\atop\kappa})$ for
the ions in more detail: Summing up the electric fields created by the charge  
fluctuations on the different layers we find:
\begin{equation}
E^\rho_z({\textstyle{n\atop\kappa}})= \cases{\frac{1}{2\epsilon_0} 
( \sum \limits_{n'\le
n} \delta\rho_{n'} -  \sum\limits_{n'>n}\delta\rho_{n'} )
& for $\kappa \in $ barrier \cr
\frac{1}{2\epsilon_0} ( \sum\limits_{n'< n} \delta\rho_{n'}- 
\sum\limits_{n'>n}\delta\rho_{n'} ) & for $\kappa \in$ layer. \cr}
\end{equation}
Note that for ions {\it inside}  the barrier the driving field 
does not depend on the position $R_z(\kappa)$ of the ion and is equal to the
constant field $E^\rho_n$ introduced in Eq.\ (\ref{Maxwell2}): 
\begin{equation}
E^\rho_z({\textstyle{n\atop \kappa}})=  \cases{E^\rho_n & for $\kappa \in$
barrier \cr 
\frac{1}{2} (E^\rho_n+E^\rho_{n-1}) & for $\kappa \in$ layer. \cr}\label{Erho} 
\end{equation}

In order to find a relation between the lattice displacements and the driving
field we introduce the Fourier transformation
\begin{equation}
E^\rho(q_z) = \sum_{n} E^\rho_n e^{-i q_z z_n}.
\end{equation}
Then 
\begin{equation}
\sum_n Z_\kappa E_z^\rho({\textstyle{n\atop\kappa}}) e^{-iq_z
z_{{n}}} =: 
\tilde Z^*_\kappa (q_z) E^\rho(q_z)
\end{equation}
with the $q_z$-dependent effective charge  
\begin{equation}
\tilde Z_\kappa(q_z) = \cases { Z_\kappa  & for
$\kappa \in$ barrier \cr \frac{1}{2}Z_\kappa 
(1+ e^{i q_z d})  &  for $\kappa \in$ layer\cr} \label{Zt} 
\end{equation} 
and consequently
\begin{eqnarray}
u_z({\textstyle{n\atop\kappa}}) &=& \frac{1}{N_z} \sum_{q_z\lambda} 
\sum_{\kappa'} 
e^{iq_zR_z({\scriptstyle{n\atop\kappa}})} \\
\nonumber
&\times&\frac {1}{\sqrt{M_\kappa}}\frac {e_z(\kappa\vert q_z\lambda)
e_z^*(\kappa'\vert q_z\lambda)}{\omega^2(q_z\lambda)-\omega^2} \frac
{e\tilde Z^*_{\kappa'}(q_z)}{\sqrt{M_{\kappa'}}}E^\rho(q_z). 
\end{eqnarray}
Thereby the vanishing of $\tilde Z_\kappa$ for $q_zd=\pi$ and for ions on the  
layers reflects the fact that for alternating electric fields in the barriers
no net force is acting on the superconducting layers.

Next we want to relate the driving field to the average total electric field $E_n$ 
in the barrier because this field is connected with the phase difference $\gamma_n$ by
the second Josephson-equation (\ref{2Jos}).
The microscopic total electric field $E_z(x,z)$ is the sum
of the fields set-up by both the charge-fluctuations and the displaced ionic
charges: $E_z(x,z) = E^\rho_z(x,z) + E^{\rm ion}_z(x,z)$. 
As $E^\rho_z(x,z)$ is assumed to be constant inside the barrier and
independent of x we may replace the other two
fields also by their averages across the barrier and over one unit cell
along the layer. For the averaged fields we have then the relation:
\begin{equation}
E^\rho_n = E_n - E^{\rm ion}_n \, . 
\end{equation}

The field $E^{\rm ion}_n$ in the barrier can be calculated from the
difference of the scalar potentials on layers $n$
and $n+1$ produced by the ionic displacements averaged over the area of 
one unit-cell as  
\begin{eqnarray}
E^{\rm ion}_n = -\frac{e}{\epsilon_0 v_c}\Bigl(&\sum\limits_{\kappa \in {\rm
barrier}}&
Z_\kappa  u_z({\textstyle{n\atop \kappa}})\\ \nonumber
 + \frac{1}{2}&\sum\limits_{\kappa \in {\rm layer}}&
Z_\kappa (u_z({\textstyle{n\atop\kappa}}) + 
u_z({\textstyle{n+1\atop\kappa}}))\Bigr).  
\end{eqnarray}  
Here $v_c$ is the volume of one unit cell. The factor $1/2$ results from the
fact that displacements of ions (with $q_\parallel =0$) on  a layer do not contribute to
the potential on the same layer. Note that this field for 
$q_\parallel=0$ does not depend on the ionic displacements in other barriers 
and  is closely related to the ionic polarisation in the same barrier. 
Defining a generalised polarisation by 
\begin{equation}
E^{\rm ion}_n=: - P_n/\epsilon_0  \label{pol},
\end{equation}
we may write for the displacement  in the barrier
\begin{equation}
D_n := \epsilon_0 E^\rho_n = \epsilon_0E_n + P_n,
\end{equation}
which has the usual form as in the macroscopic Maxwell theory.

Going over to a Fourier transformation  the relation 
\begin{equation} 
P(q_z) = \chi(q_z,\omega) \epsilon_0E^\rho(q_z) 
\end{equation}
between the
polarisation and the driving field is obtained with 
\begin{equation} 
\chi(q_z,\omega) = \sum_\lambda \frac {\vert\Omega(q_z\lambda)\vert^2}
{\omega^2(q_z \lambda) - \omega^2} \label{chi} 
\end{equation} 
and the oscillator strength 
\begin{equation} 
\vert\Omega(q_z\lambda)\vert^2 =
\frac{e^2}{v_c\epsilon_0} \sum_{\kappa\kappa'} \tilde Z_\kappa(q_z)
\frac {e_z(\kappa\vert q_z\lambda) e^*_z(\kappa'\vert q_z\lambda)}
{\sqrt{M_\kappa M_{\kappa'}}}\tilde Z^*_{\kappa'}(q_z). 
\label{osc} \end{equation} 
The special combination of phase factors contained in $\tilde Z_\kappa(q_z)$
Eq.\ (\ref{Zt}) are 
a consequence of the different contribution of ions on and between the
superconducting layers to the electric field in the barrier. 

Using $\epsilon_0 E^\rho(q_z) = \epsilon_0 E(q_z) + P(q_z)$ we
can solve for $P(q_z)$: 
\begin{equation}
P(q_z) = \frac{\chi(q_z, \omega )}{1-\chi(q_z, \omega )} \epsilon_0
E(q_z ).
\end{equation}
The relation $D(q_z) = \epsilon_0 \epsilon^L_{\rm ph}(q_z, \omega) 
E(q_z )$   defines an effective longitudinal dielectric function 
\begin{equation} 
\epsilon^L_{\rm ph}(q_z, \omega) = \frac{1}{1-\chi(q_z, \omega)}  \label{ephon} 
\end{equation}
This function has zeros at the eigenfrequencies $\omega( q_z \lambda)$ of the
dynamical matrix. Due to the form of the oscillator strengths 
Eq.\ (\ref{osc}) only modes with
polarization in the c-direction contribute. As the electric field $E_n$ is
constant  along the layers we have $\vec q_\parallel=0$ and only
longitudinal modes in the c-direction couple, therefore the zeros of 
$\epsilon^L_{\rm ph}(q_z,\omega)$ are exactly at the longitudinal
eigenfrequencies of the dynamical matrix. 

In the case of a single dispersionless phonon-mode with frequency $\omega_L$ 
the function $\epsilon^L_{\rm ph}(\omega)$ can be directly compared with the 
dielectric function used in Ref.\ \onlinecite{wir1}. In fact, in this
case  $\epsilon^L_{\rm ph}(\omega)$ can be written as  
\begin{equation}
\epsilon^L_{\rm ph}(\omega) = 1 + \frac {\Omega^2}{\omega_L^2 - \Omega^2 -
\omega^2}. \label{eps}
\end{equation}

The form of the  longitudinal dielectric function 
$\epsilon^L_{\rm ph}(q_z,\omega)=\epsilon^{zz}_{\rm ph}(q_z,\omega))$
Eq.\ (\ref{ephon}), which we have introduced here, is 
different from the transverse dielectric function $\epsilon^T_{\rm
ph}(q_x,\omega)=\epsilon^{zz}_{\rm ph}(q_x,\omega))=1+\chi(q_x,\omega)$ 
In both functions different eigenfrequencies and
oscillator strengths enter, however in the limit $q_z\to 0, q_x \to 0$, which 
is only relevant in optical experiments,  the values of both functions are equal.

Finally a comparison of our theory with theoretical investigations in Ref.\ 
\onlinecite{Maksimov} are in order. In principle there are two
different electron-phonon coupling mechanisms, which may couple Josephson
oscillations and phonons: 1. the electromagnetic interaction between the ionic
charges and the charges of conduction electrons, 2. the dependence of the
tunneling matrix element on lattice displacements. The first mechanism is
considered in our work,  the second in Ref.\ \onlinecite{Maksimov}. Both 
mechanisms require a
different theoretical treatment (on a diagrammatical basis the two mechanisms
would correspond to different diagrams).  It has been argued in 
Ref.\ \onlinecite{Zeyher} 
that in the layered cuprate superconductors the charges  of the
ions in the insulating barrier between superconducting layers are unscreened   
and therefore have a strong interaction with conduction electrons in the
CuO$_2$-layers. We therefore considered this mechanism for our treatment of the
coupling between Josephson oscillations  and phonons. Though we did not write 
down a Hamiltonian for the interacting system our method
nevertheless  
is a full microscopic theory which treats the
electron-phonon interaction on a random-phase-type level by describing the
interaction with the help of internal fields. This approximation is sufficient
as long as we do not want to consider the electron-phonon
interaction inside the superconducting layers and treat exchange
effects between different superconducting layers.

 \section{Influence of phonons on the IV-characteristic}  

According to the RSJ-like model derived in Eq.\ (\ref{RSJ0}) the current 
density in
junction $n$ is 
\begin{equation}
j= j_c \sin \gamma_n(t) + \sigma E_n(t) +  \dot D_n(t) \label{RSJ1}
\end{equation}
where $D_n(t)=\epsilon_0 E^\rho_n(t)$ is the electric field in junction $n$
set-up by the charge fluctuations of conduction electrons.  As pointed
out in the previous section  this field can be expressed by
the average electric field in the barrier and the generalized polarisation     
Eq.\ (\ref{pol}) as  $D_n(t) = \epsilon_0 E_n(t) +
P_n(t)$.
The polarisation has to be calculated self-consistently from 
the ionic displacements and depends linearly on the electric field. 

Let us discuss first the case of one resistive junction at $n=0$ in the
middle of a large stack while
all other junctions $n\ne 0$ are in the superconducting
state. Then as mentioned previously all the oscillations are governed by one    
frequency $\omega$, 
and  we can write for the  phase for $n=0$: 
\begin{equation} 
\gamma_0(t)= \theta_0 + \omega t + \delta \gamma_0(t),
\end{equation}
while for $n\ne 0$ we have 
\begin{equation}
\gamma_n(t) = \theta_n + \delta\gamma_n(t).
\end{equation}
In the stationary state $\delta \gamma_n(t)$ oscillates with the same frequency
$\omega$, 
\begin{equation}
\delta \gamma_n(t)=  \delta\gamma_n e^{-i\omega t} + c.c.
\end{equation}
Higher harmonics can be neglected for $\beta_c\gg 1$. 
In this limit the fluctuations 
$\delta \gamma_n(t)$ are small and we may use the expansion     
\begin{equation} 
\sin \gamma_0(t) \simeq \sin(\theta_0 + \omega t) + \cos(\theta_0 + 
\omega t) \delta \gamma_0(t),
\end{equation}
while for $n \ne 0$ we have
\begin{equation} 
\sin \gamma_n(t) \simeq \sin\theta_n + \cos\theta_n \delta
\gamma_n(t). 
\end{equation}

The bias current density $j$ on the l.h.s. of the RSJ-equations 
Eq.\ (\ref{RSJ1}) is
time-independent and equal for all junctions, while the quantities on the
r.h.s have both time-independent and oscillating components.

Let us discuss  the equations for the non-resistive junctions ($n\ne 0$) first.
Here the dc-component is:
\begin{equation} 
j=j_c\sin \theta_n \, .
\end{equation}
This fixes the constant part of the phases in the non-resistive junctions and
relates it to the  bias current.

For the oscillating part of Eq.\ (\ref{RSJ1}) one obtains:
\begin{equation}
0=j_c\cos \theta_n \delta \gamma_n(t) + \sigma \frac{\hbar}{2ed} \delta
\dot \gamma_n(t) + \dot D_n(t) \label{nne0}
\end{equation}
or
\begin{equation}
0=\bar \omega_p^2 \delta \gamma_n(t) + \frac{\sigma}{\epsilon_0}
\delta \dot \gamma_n(t) + \frac{2ed}{\hbar\epsilon_0} \dot D_n(t)
\end{equation}
with the reduced Josephson plasma frequency
\begin{equation}
\bar \omega_p^2= \omega_p^2\sqrt{1- \left( \frac{j}{j_c} \right)^2}. 
\end{equation}

Now we discuss the resistive junction at $n=0$. 
Keeping only the lowest harmonic we find
\begin{equation}
\sin \gamma_0(t) \simeq  \sin(\theta_0 + \omega t) + \Re
(\delta \gamma_0 e^{i\theta_0}).
\end{equation} 
The dc-component of the RSJ-equation Eq.\ (\ref{RSJ1}) is therefore given by
\begin{equation}
j(V)=j_{\rm qp}(V) + j_c \Re (\delta\gamma_0 e^{i\theta_0}),
\end{equation}
where $V$ is the dc-voltage of the 
resistive junction and $j_{\rm qp}(V)=\sigma E_{dc}$ is the quasiparticle
current density. 

For the oscillating part
one finds: 
\begin{equation}
0=j_c\sin(\theta_0 + \omega t) + \sigma
\frac{\hbar}{2ed} \delta \dot \gamma_0(t) + 
\dot D_0(t). \label{ne0}
\end{equation}

The two equations (\ref{nne0}, \ref{ne0}) can be combined to one  inhomogeneous
linear differential equation for all $n$
\begin{equation}
\bar \omega_p^2 \delta \gamma_n(t) + \frac{\sigma}{\epsilon_0}
\delta \dot \gamma_n(t) + \frac{2ed}{\hbar\epsilon_0} \dot D_n(t) = f_n(t)
\end{equation}
with
\begin{equation} 
f_n(t)= \cases {\bar\omega_p^2\delta \gamma_0(t) -  \omega_p^2 \sin
(\theta_0 + \omega t) & for $n=0$ \cr
0 & for $n \ne 0$. \cr}
\end{equation}

Assuming a time dependence of the form $e^{-i\omega t}$ for all oscillating
quantities we have
\begin{equation}
\bar \omega_p^2 \delta \gamma_n + \frac{-i\omega\sigma}{\epsilon_0}
\delta \gamma_n +
\frac{2ed}{\hbar\epsilon_0} (-i\omega)D_n = f_n \label{inh}
\end{equation}
with 
\begin{equation}
f_n= \cases{\bar \omega_p^2 \delta \gamma_0 -
\frac{i\omega_p^2}{2}e^{-i\theta_0} & for $n=0$ \cr
0 & for $n \ne 0$. \cr}
\end{equation}

In order to incorporate the non-local dependence of the polarisation on the
electric fields in different barriers a spatial Fourier
representation of the form  
\begin{equation}
\delta \gamma_n = \frac{1}{N_z}\sum_{q_z} \gamma(q_z) e^{iq_z z_n}
\end{equation}
is introduced. Using the relation  
\begin{equation}
D(q_z)= \epsilon_0\epsilon^L_{\rm ph}(q_z,\omega)E(q_z) = 
\frac{\epsilon_0\hbar}{2ed}\epsilon^L_{\rm ph}(q_z,\omega)(-i\omega)            
\gamma(q_z)
\end{equation}
in Eq.\ (\ref{inh}) yields
\begin{equation}
G^{-1}(q_z,\omega) \gamma(q_z) = f_0
\end{equation}
with
\begin{equation}
G^{-1}(q_z,\omega) = \bar\omega^2_p - i\omega \frac{\sigma}{\epsilon_0} -
\omega^2 \epsilon^L_{\rm ph}(q_z,\omega). 
 \end{equation}
For the phase oscillation in the resistive junction follows 
\begin{equation}
\delta\gamma_0 = \frac{1}{N_z}
\sum_{q_z} \gamma(q_z) = g(\omega) f_0
\end{equation}
 with
\begin{equation}
g(\omega )= \frac{1}{N_z}\sum_{q_z} G(q_z, \omega).
\end{equation}
Solving for $\delta \gamma_0$ 
\begin{equation}
\delta \gamma_0= \frac{1}{2}\frac{-i\omega_p^2}{g^{-1}(\omega) - \bar \omega
^2_p}e^{-i\theta_0} 
\end{equation}
 we obtain
\begin{equation}
\Re (\delta \gamma_0 e^{i\theta_0}) = \frac{1}{2} \Im \frac
{\omega_p^2}{g^{-1}(\omega) - \bar \omega_p^2}. 
\end{equation}
From this finally the following expression for the dc-current density as
function of the dc voltage is obtained: 
\begin{eqnarray}\label{IV1}
j(V)&=& j_{\rm qp}(V) - \frac{j_c}{2} \frac{\omega^2_p}{\omega^2} \Im \frac
{1}{\tilde \epsilon(\omega)}\\ \nonumber
&=& j_{\rm qp}(V) + \frac{j_c}{2} \frac{\omega^2_p}{\omega^2} \frac
{\tilde\epsilon_2(\omega)}{\tilde \epsilon_1^2(\omega) + \tilde
\epsilon_2^2(\omega)}.  
\end{eqnarray}
Here $\tilde\epsilon(\omega)$ is a modified dielectric function
\begin{equation}
\tilde \epsilon(\omega) = \bar \epsilon_{\rm ph}(\omega) +
\frac{i\sigma}{\epsilon_0\omega} 
\end{equation}
where 
\begin{equation}
\bar \epsilon_{\rm ph}(\omega)= J^{-1}(\omega) + 
\frac{\bar\omega^2_p}{\omega^2} -
\frac{i\sigma}{\epsilon_0 \omega} \label{bareps}
\end{equation}
and
\begin{equation}
J(\omega) = \frac{1}{N_z} \sum_{q_z}
\bigl[\epsilon^L_{\rm ph}(q_z,\omega) - \frac{\bar\omega^2_p}{\omega^2} +
\frac{i\sigma}{\epsilon_0 \omega}\bigr]^{-1}. \label{J}
\end{equation}

\begin{figure}
\epsfig{figure=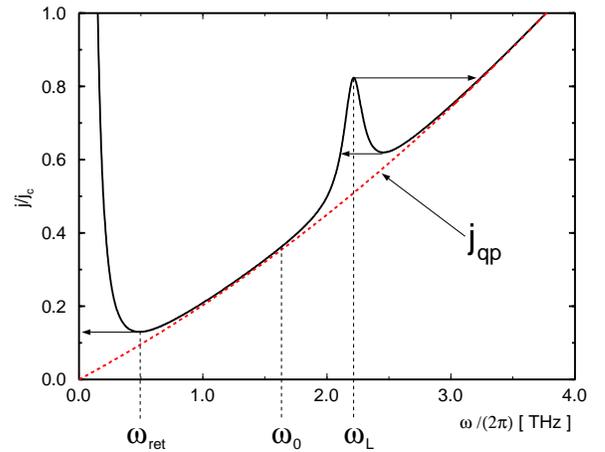,height=6cm}
\caption{ Analytical $I$-$V$-curve for one resistive junction 
with one
phonon resonance at $\omega=\omega_L$. The arrows mark the hysteretic jumps
found in  current-biased experiments and numerical simulations. \label{fig2}}
\end{figure}

This expression describes the dc-current density as function of the dc-voltage 
$V= \hbar \omega/(2e)$. It has a maximum for frequencies $\omega$ where
the real part $\tilde \epsilon(\omega)$ vanishes, i.e. for longitudinal
phonon frequencies. 

This can  easily be seen if we consider the special case of one
phonon mode without dispersion. Then $\bar \epsilon_{\rm ph}(\omega)=
\epsilon^L_{\rm ph}(\omega)$, which  is of
the form Eq.\ (\ref{eps}). 
The corresponding $I$-$V$-curve is shown in Fig.\ \ref{fig2}, which is
calculated with the dielectric function
\begin{equation}
\epsilon^L_{\rm ph}(\omega) = 1 + \frac {\Omega^2}{\omega_L^2 - \Omega^2 -
\omega^2 - i\omega \rho}. 
\end{equation}
Here an additional 
damping $\rho$ of the phonon has been introduced in Eq.\ (\ref{eps}) in order   
to simulate the energy transfer into other junctions due the coupling of 
the ions. A peak in the
$I$-$V$-curve appears at
$\omega = \omega_L$. The width of the peak is determined by this damping 
together with the
quasiparticle conductivity. The deviation from the quasiparticle current 
density
vanishes at the pole of $\epsilon^L_{\rm ph}(\omega)$ at
$\omega_0=\sqrt{\omega^2_L - \Omega^2}$.   
The rise at low voltages is due to the plasma
resonance. For $\beta_c \gg 1$ the minimum of the $I$-$V$-curve is at        
$\omega_{ret}\simeq (3/2)^{1/4}\omega_p/\sqrt{\epsilon^L_{\rm ph}(\omega_p)}$.
 In 
current-biased  experiments and corresponding numerical simulations the
parts with negative differential conductivity are  skipped hysteretically as   
indicated in the figure.

Note that the denominator in the function $J(\omega)$ Eq.\ (\ref{J}) is the 
total
$q_z$- and $\omega$-dependent longitudinal dielectric function of the          
coupled system of phonons and conduction electrons, 
\begin{equation}
\epsilon^L_{\rm tot}(q_z,\omega) = \epsilon^L_{\rm ph}(q_z,\omega) - 
\frac{\bar\omega^2_p}{\omega^2} +
\frac{i\sigma}{\epsilon_0 \omega}. \label{epsL}
\end{equation}
Zeros of the real part of this function describe longitudinal collective       
modes in the system. On the other hand the resonances in the $I$-$V$ curve
appear at the {\it bare} longitudinal phonon frequency in the case of a narrow
phonon band. The  summation over $q_z$ in Eq.\ (\ref{J}) leads to an
effective damping of the resonances which is proportional to the phonon
dispersion. The physical origin is the loss of energy by phonons from the
resistive junction to the neighboring junctions.

The result for the current-voltage characteristic can be generalised to the
case of several junctions being in the resistive state, if we assume that
all junctions oscillate with the same frequency $\omega$. Denoting the
subset of indices for the resistive junctions  by $I$ then for $i \in I$ 
we obtain (for a derivation see the appendix):
\begin{eqnarray}\label{multi1} 
j(V)&=& j_{\rm qp}(\frac{\hbar \omega}{2e}) \\ \nonumber
&-&\frac{j_c}{2}\frac{\omega^2_p}{\omega^2} \Im \sum_{k \in I} e^{i\theta_{i}}
\Big[ \bar \epsilon(i,k,\omega) +
\frac{i\sigma}{\epsilon_0\omega}\delta_{i,k}\Bigr]^{-1}
e^{-i\theta_{k}}  
\end{eqnarray} 
The dielectric function $\bar \epsilon(i,k,\omega)$ is defined by 
\begin{eqnarray}\label{multi2}
\bar \epsilon(i,k,\omega) &:=& \Bigl[\frac{1}{N_z} \sum_{q_z}\frac{e^{iq_z(z_i-
z_k)}} {\epsilon^L_{\rm ph}(q_z,\omega)
- \frac{\bar\omega^2_p}{\omega^2} +
\frac{i\sigma}{\epsilon_0 \omega}}\Bigr]^{-1} \\ \nonumber
&+&(\frac{\bar\omega^2_p}{\omega^2} -
\frac{i\sigma}{\epsilon_0 \omega})\delta_{i,k} 
\end{eqnarray}

The terms in 
brackets in Eqs.\ (\ref{multi1}, \ref{multi2}) are understood as matrix
inversions. The dc-voltage $V$ is obtained, if one multiplies 
$\hbar\omega/(2e)$
by the number of resistive junctions. Note that the r.h.s. of 
Eq.\ (\ref{multi1}) 
depends on the layer index $i \in I$, while the l.h.s. is equal 
for  each junction. From this equality the (relative) phases $\theta_i$ in the
different    junctions can in principle be determined, which in turn provides 
an analytical expression for the $I$-$V$-curve. 

In the case of two resistive junctions exactly two solutions exist with
$\theta_i=\theta_j$ and $\theta_i=\theta_j+\pi$, respectively. In the general
case several different solutions are found. The stability of these
solutions will be checked by a comparison with a direct numerical integration
of the coupled equations of motions in the following section. 
It turns out, that those analytical solutions are most stable where the
phases $\theta_i$ of the oscillating Josephson junctions fit best to
the pattern of lattice vibrations with the given frequency $\omega$. 

\section{A simple example}

The theory in the preceding sections is developed for  general
lattice dynamical models and is valid within the assumptions we have made for
the superconducting properties: we treat only single-layer
systems assuming a homogeneous conduction electron charge distribution
along the layers. An extension of the theory to more realistic systems is in 
principle
possible, but requires to introduce the charge susceptibility of conduction
electrons in the superconducting state and a generalisation of the Josephson
theory to multilayer systems. 
But for this more details of the electronic properties are required than  
are currently known about these materials. 
In addition to this, it is not possible to compare theoretical 
results for the $I$-$V$-curves with experiments in detail,
as reliable lattice dynamical calculations for the strongly anisotropic 
 systems BSCCO and TBCCO with two and three layers and variable oxigen content
 are not yet available. 

In the following we consequently only  want to illustrate the main features 
of our theory in a simple toy model, which reflects some basic aspects of the 
real system. 
 One of the main lattice dynamical property of
these systems is certainly the existence of a longitudinal acoustical and
(several) flat longitudinal optical bands which result from movements of
groups of ions in the barrier against ions in the
CuO$_2$-planes in the c-direction.\cite{kulkarnitl1,kulkarnitl2,pradeBi1}   
Such modes we simulate by the most simple lattice dynamical model
consisting of 
two kinds of  ions with ionic charges $Z_l$, $Z_b=-Z_l$ and masses
$M_l$, $M_b$. The first kind ($\kappa=l$) is placed on the  superconducting
layers, the second kind ($\kappa=b$) in the middle of the barrier. The motion
of ions in the c-direction which is assumed to be uniform along the layers is
approximated  by a two-atomic chain-model with nearest-neighbor interactions in
the c-direction:  \begin{eqnarray}\label{vib}   M_l\ddot u({\textstyle{n\atop l}})
- f(u({\textstyle{n \atop b}}) + u({\textstyle{n-1\atop b}}) - 2
u({\textstyle{n \atop l}})) &=& Z_l E^\rho({\textstyle{n \atop l}})  \cr 
M_b\ddot u({\textstyle{n \atop b}}) - f(u({\textstyle{n+1\atop l}}) +
u({\textstyle{n\atop l}}) - 2 u({\textstyle{n\atop b}})) &=& Z_b
E^\rho({\textstyle{n\atop b}}).     
\end{eqnarray}
By choosing the masses very different a
narrow optical band can be simulated. 
From a diagonalisation of the dynamical matrix given by Eq.\ (\ref{vib}) the 
well-known
eigenfrequencies $\omega(q_z\lambda)$ of the two-atomic chain are obtained.    
With help of the eigenvectors the oscillator strengths 
defined in Eq.\ (\ref{osc}) are calculated, which are needed for the 
longitudinal
dielectric function Eq.\ (\ref{ephon}).

The driving field on the r.h.s. of Eq.\ (\ref{vib}) is the field set-up by the  
conduction  electron charges on
the superconducting layers, which  can be expressed by the (constant) field
$E^\rho_n$ in the barrier between layers $n$ and $n+1$ with the help of
Eq.\ (\ref{Erho}). 
The latter can be expressed by the average total electric field $E_n$ in
the barrier: 
\begin{equation}
E^\rho_n = E_n + \frac{1}{\epsilon_0} P_n
\end{equation}
where the polarisation Eq.\ (\ref{pol}) is given by:
\begin{equation}
P_n= \frac{e}{v_c} \Bigl( Z_b u({\textstyle{n\atop b}}) +
\frac{1}{2}Z_l\bigl( u({\textstyle{n\atop l}}) + 
u({\textstyle{n+1\atop l}})\bigr)\Bigr).
\label{pol1}  
\end{equation}

These equations for the motion of lattice displacements have to be
supplemented by the extended RSJ-equations Eq.\ (\ref{RSJ1}):
\begin{equation}
j = j_c \sin \gamma_n(t) + \sigma \frac{\hbar}{2ed} \dot \gamma_n(t)
+ \epsilon_0 \frac {\hbar}{2ed }\ddot \gamma_n(t) + \dot P_n(t). \label{RSJ}
\end{equation}

This model is used to calculate the $I$-$V$-characteristics in two ways: 1.
The $I$-$V$-curves are calculated analytically using the results
Eq.\ (\ref{multi1}) obtained by the Green's function method. Thus the peaks due 
to the phonon resonances
are obtained. 2. The coupled set of RSJ equations Eq.\ (\ref{RSJ}) and phonon 
equations
Eqs.\ (\ref{vib}, \ref{pol1}) are integrated numerically by a Runge-Kutta 
method 
for a finite stack of Josephson junctions. Changing the bias-current gradually  
allows  to follow  the $I$-$V$-curves as in the current-biased experimental 
situation
and to reproduce the hysteretic behaviour in particular.

We start with the 
discussion of the first branch of  the $I$-$V$-curve, where one junction is in 
the resistive state.

\subsection{One resistive junction}

Quite generally the $I$-$V$-curve is expected to have peaks at the van Hove
singularities of the phonon dispersion.  
Details, however, depend on the
oscillator strength defined by Eq.\ (\ref{osc}) which enters the longitudinal
dielectric function Eq.\ (\ref{ephon}). In particular at the edge of the 
Brillouin
zone for $q_z= \pi/d$ only the motion of ions within the barrier contribute to 
the oscillator strength, the ions on the superconducting layers are inactive
due to the factor $1 + \exp(iq_zd)$ in Eq.\ (\ref{Zt}). These features will be  
illustrated in the following.  

\begin{figure}
\epsfig{figure=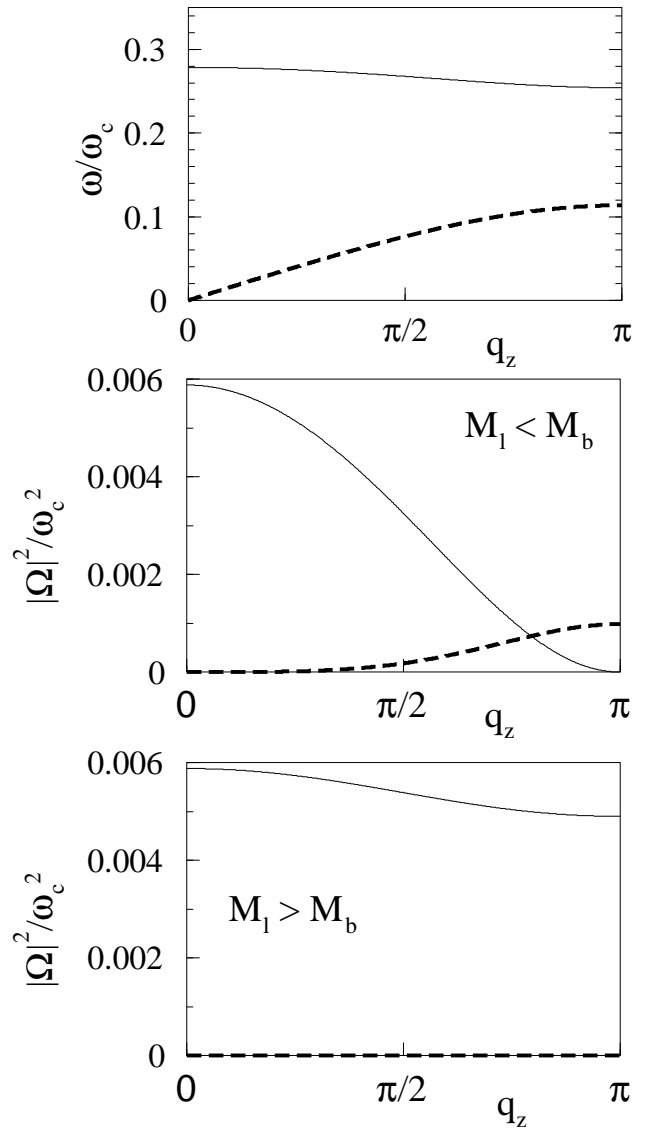, height=15cm}
\caption{ Dispersion and oscillator strengths for the two-atomic
chain model. Shown are the oscillator strengths   
$\vert\Omega(q_z\lambda)\vert^2$ for the acoustical branch (dashed curve) 
and the optical branch (solid curve) and for the two cases of the heavy ions    
on the superconducting layer ($M_l>M_b$) and in the barrier ($M_l<M_b$). 
\label{fig3}}  
\end{figure}  
For the lattice
dynamical model introduced above at $q_z= \pi/d$  only
one type of particles is moving due to symmetry: In the 
acoustical branch the heavier ion, in the
optical branch  the lighter ion is moving. If  the heavier ion is 
on the superconducting layers ($M_l > M_b$) the oscillator strength vanishes      
at the end of the acoustic branch (see Fig.\ \ref{fig3}), and peaks are 
expected to appear in the
$I$-$V$-curve at the two van-Hove singularities of the optical branch. On the
other hand, if the lighter ion is on the superconducting layers ($M_l<M_b$)
then the oscillator strength vanishes at the end of the optical branch, and 
peaks are expected at $q_z=\pi/d$ from the acoustical branch and at $q_z=0$
from the optical branch.

\begin{figure}
\epsfig{figure=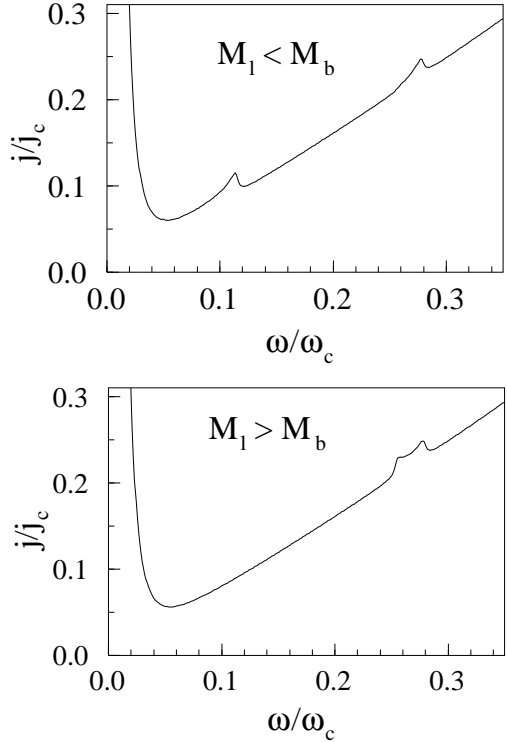, height=10cm}
\caption{ $I$-$V$-curves for one resistive junction with 
subgap-structures
due to acoustical and optical phonons. \label{fig4}}
\end{figure}  

\begin{figure}
\epsfig{figure=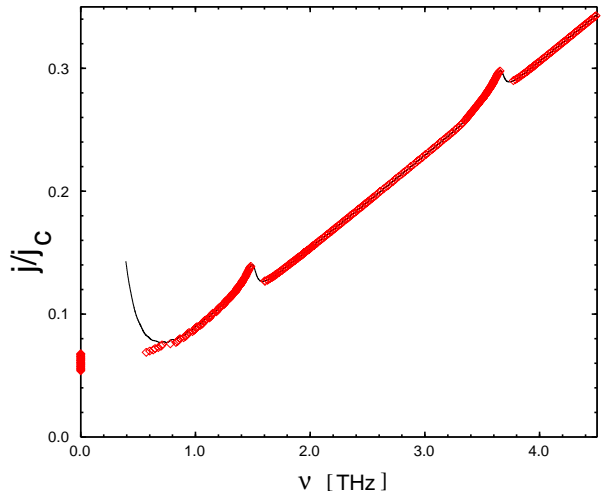, height=6.5cm}
\caption{ Comparison between analytical (solid line) and 
numerical results (rhombs) for the $I$-$V$-curve of  one resistive junction 
with subgap-structures
due to acoustical and optical phonons for $M_l <M_b$. \label{fig5}}
\end{figure}

This is illustrated in Fig.\ \ref{fig4}  where we have plotted  
results for the $I$-$V$-curve 
of the first branch in the two cases $M_l<M_b$ and $M_l>M_b$. In our model the 
phonon
dispersion is fixed by the values of  $\omega^2_{\rm LO}(q_z=0)=
2f(1/M_l+ 1/M_b)$ and the mass ratio $M_l/M_b$. A measure for the oscillator
strength is the quantity $\Omega_{l,b}^2:= Z^2/(M_{l,b} \epsilon_0 v_c)$. In 
Fig.\ \ref{fig4} we
have used the following parameters: $\beta_c= \omega_c^2/\omega_J^2= 375$,
$M_l/M_b = 0.2$ and $M_b/M_l=0.2$ respectively.
 The phonon frequencies are normalised to $\omega_c$ and are given by 
$\omega_{\rm LO}(q_z=0)= 0.28 \omega_c$. The oscillator strength is
given by $\omega_c^2/\Omega_l^2=200 $ for $M_l<M_b$ and
$\omega_c^2/\Omega_b^2=200 $ for
$M_l>M_b$. The corresponding phonon dispersion and the oscillator
strengths are shown in Fig.\ \ref{fig3}. The values of the parameters
$\beta_c$, $\omega_{LO}/\omega_c$ are adapted to TBCCO, also the chosen value
$|\Omega|^2 / \omega_0^2 =0.08$ for the oscillator strength 
compares well to the experimental estimate $\approx 0.13$ \cite{tsvetkov}. 
The mass-ratio is chosen to obtain a sufficiently flat optical branch 
(cf. the band structure in Ref. \onlinecite{kulkarnitl2}).

For $M_l<M_b$ the analytical results for the $I$-$V$-curve in Fig.\
\ref{fig4}
show phonon peaks at the van Hove singularity of the acoustic branch at
$q_z= \pi/d$ and at the van Hove singularity at $q_z=0$ of the optical branch,
the resonance at the van Hove singularity at $q_z=\pi/d$ of the optical
branch is suppressed. For $M_l>M_b$ only structures due to the two
van Hove singularities of the optical branch appear. In both cases the
increase of the $I$-$V$-curve at low frequencies indicates the  Josephson
plasma frequency. The numerical results shown in Fig.\ \ref{fig5} for
$M_l<M_b$ (here $\omega_c$ = 13.1 THz) follow the analytical results  in the
regions of positive differential resistance perfectly, and otherwise show the
hysteretic behaviour as seen in experiments.  At low values of $j/j_c$ the
$I$-$V$-curve switches back to the superconducing state of the junction. 

\subsection{Two resistive junctions}

Another important effect of the coupling between Josephson oscillations
and phonons is the synchronisation (phase-locking) of Josephson oscillations in
different resistive junctions, which would be absent without phonons 
in short junctions, which are homogeneous parallel to the layers.
we want to illustrate this for
the case of two resistive junctions coupled by one narrow optical phonon branch. 

\begin{figure}
\epsfig{figure=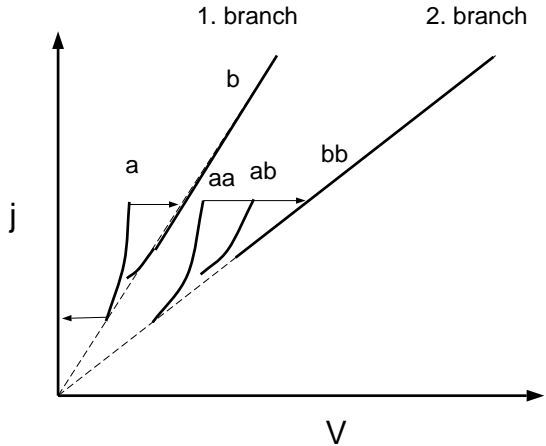, height=6cm}
\caption{ Schematic plot of the first and second branch of the
$I$-$V$-curve with subgap structures due to one phonon. \label{fig6}}
\end{figure}

The second branch of the $I$-$V$-curve which corresponds to two resisitve
junctions has  a rather complex structure already  for one
phonon band. This is shown schematically in Fig.\ \ref{fig6}. If we denote the
two dynamical states of the first branch by $a$ and $b$, then on the second
branch with two resistive junctions either both junctions are in state $a$
(label $aa$),  both junctions are in state $b$ (label $bb$) or one junction is
in state $a$ while the other junction is in state $b$ (label $ab$). Note that
in the latter case the oscillation frequencies of the two junctions are
different.  In the case of well separated resistive junctions the voltages of
a structure for a given bias current are determined by $\omega_{aa}=
2\omega_a$, $\omega_{bb}= 2\omega_b$, $\omega_{ab}=\omega_a+\omega_b$. This is
no longer true, if the resistive junctions are close to each other and
interact via phonons. Then the voltages in the second branch are slightly
lower.

In the case of two resistive junctions $i$ and $j$ two solutions of
Eq.\ (\ref{multi1}) exist with phase differences 
$\theta_i=\theta_j$ and $\theta_i=\theta_j+
\pi$, corresponding to in-phase and out-of-phase Josephson oscillations,
respectively. Inserting these results in Eq.\ (\ref{multi1}) two different  
$I$-$V$-curves can be calculated. 
Note that this formula applies only for the states
$aa$ and $bb$, because in the derivation we have assumed that the two
junctions oscillate with the same frequency. A similar formula can also be
derived for the state $ab$. In that case the phases are meaningless 
because the two junctions  oscillate with different
frequencies and are essentially decoupled. 

\begin{figure}
\epsfig{figure=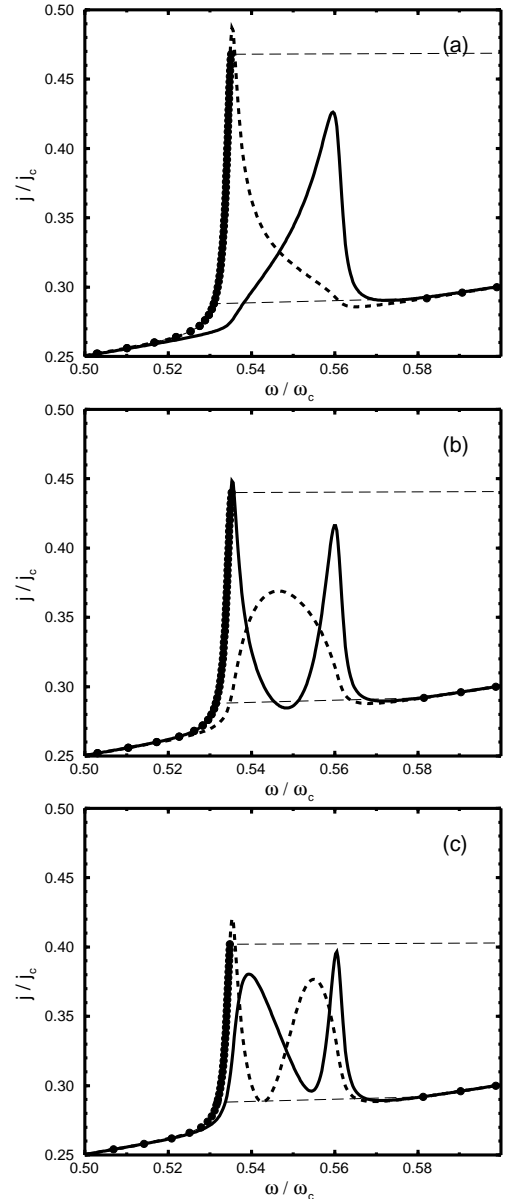, height=16cm}
\caption{ $I$-$V$-curve for two
resistive junctions separated by 0 (a), 1 (b), 2  (c) superconducting
junctions. Shown are analytical $I$-$V$-curves calculated for a narrow
optical band for the in-phase, $\theta_j -\theta_i=0$ (solid line) and the 
out-of-phase, 
$\theta_j- \theta_i=\pi$ (dashed line) 
solution together with numerical results (dots). \label{fig7}} 
\end{figure}

In Fig.\ \ref{fig7} we show examples calculated for a narrow
optical band with $M_l/M_b=10$ and the light ion oscillating in the barrier.
Here
the different analytical solutions are shown together with numerical results
for neighboring resistive junctions, $j=i+1$ (a),  and two resistive junctions 
separated by 1 or 2 superconducting junctions, $j=i+2$ (b), $j=i+3$ (c). 

It is plausible  that in the case of neighboring 
resistive junctions out-of-phase Josephson oscillations 
($\theta_i = \theta_j + \pi$)
favor a coupling to phonons at the edge $q_z = \pi / d$
of the Brillouin zone, while the coupling of in-phase
oscillations ($\theta_i = \theta_j$)
is strongest for zone-center phonons at $q_z=0$.
 This is shown in Fig.
\ref{fig7}a, where the $I$-$V$-curve for the in-phase solution shows a
peak at $\omega(q_z=0)$, while for the out-of-phase solution the
current-maximum is at $\omega(q_z=\pi/d)$. 

It can be seen that the numerical results in the dynamical state $aa$  follow 
one of
the analytical solutions before a hysteretic switch into state $bb$ occurs
(outside the figure). This is verified in Fig.\ 
\ref{fig7}a where the numerical $I$-$V$-curve follows the analytical curve
for  $\theta_i=\theta_{i+1}+\pi$. In Fig.\ \ref{fig7}b the in-phase solution 
with
$\theta_i=\theta_{i+2}$ has maxima for voltages  corresponding to frequencies
of optical phonons  at both $q_z=\pi/d$ and $q_z=0$, while the out-of-phase
solution has a broad maximum in the middle of the Brillouin zone. The
numerical results follow the in-phase solution.  Fig.\ \ref{fig7}c shows
results for $j= i+3$. 

\subsection{Several resistive junctions}

The results obtained for two resistive junctions show that the peak
position in the $I$-$V$-curve depends only slightly on the distance between
the resistive junctions. More important is the fact that 
phonons are able
to synchronise the phases of Josephson oscillations in different resistive
junctions. This is important for the use of such systems in high-frequency
applications. We have checked this numerically for the case of a block of many
resistive junctions. For frequencies $\omega$ close to a phonon eigenfrequency 
$\omega(q_z=\pi/d)$ at the edge of the Brioullin zone
the Josephson field oscillations in neighboring junctions differ by
a value close to $\pi$. For frequencies close to $\omega (q_z=0)$ near
the $\Gamma$-point the
Josephson oscillations are nearly in phase. These phase-locked dynamical
states are reached from arbitrary initial conditions for the phases. There is also a
synchronisation of Josephson oscillations for frequencies far away from phonon
resonances. We did not yet investigate the stability of these states
systematically but we expect these to be less stable than at frequencies close to a phonon
resonance.

\section{Experimental results}

Recently the explanation of the subgap resonances in 
Refs.\ \onlinecite{schlengal,yurgenss,seidel,wir3} with  the 
phonon coupling mechanism presented here could be well confirmed by 
Raman measurements on the same samples \cite{yurgenss,yurgenspriv}
and infrared reflectivity experiments with grazing 
incidence,\cite{tsvetkov,tajima} where the latter allows to determine both 
longitudinal and 
transverse modes (see Table \ref{phonon_tab}). 
Small deviations of the order of $\sim 5-10 \%$ may be attributed to the 
fact that in optical experiments and in the intrinsic Josephson effect
different averages over $\vec q$ of the dielectric functions are relevant.
Note that in our theory  modes which are 
Raman active at $q_z=0$ may
couple to intrinsic Josephson oscillations also for $q_z\ne 0$.
Earlier experimental data,\cite{Zett,Buck} which are obtained from  
polycrystalline samples, show the same qualitative behaviour,
but differ in detail.

With the help of rigid-ion 
\cite{jia} and shell-model calculations
\cite{kulkarnitl1,kulkarnitl2,pradeBi1} some of the more pronounced 
structures can be connected with certain elongation patterns of the ions in 
the unit cell. 
For example, the peak in the $I$-$V$-curve at 4.64 THz in 
Tl$_2$Ba$_2$Ca$_2$Cu$_3$O$_{10}$ seems to be due to a (Cu,Ba)-mode.

The qualitative features of the subgap resonances  have already been 
explained with the help of the phonon interpretation in 
Ref.\ \onlinecite{wir3}: The 
position of the resonance  is independent  on temperature, magnetic 
field and the geometry of the probe, while the intensity of the structure 
varies $\sim j_c^2$ with the critical current density $j_c(T,B)$. 
Also the behaviour of the position and intensity of the  structures in 
external pressure are in agreement with the phonon explanation and formula 
(\ref{IV-curve}).

One of the main qualitative features of the general theory with dispersive 
phonon bands, which goes beyond the local oscillator model used 
in Refs.\ \onlinecite{wir1,wir3}, is the possibility to describe resonances at 
van-Hove singularities, which appear e.g. at the upper band edge of the 
acoustical phonon band. 

\begin{figure}
\epsfig{figure=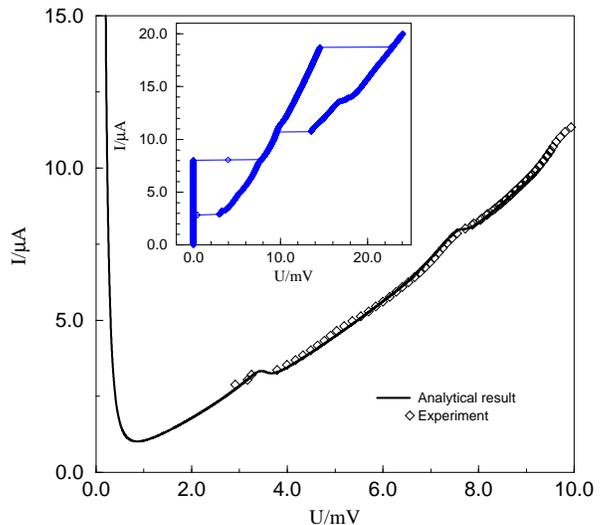, height=8cm}
\caption{ Fit of the 
experimental $I$-$V$-curve in 
Tl$_2$Ba$_2$Ca$_2$Cu$_3$O$_{10}$ from Ref.\ \protect\onlinecite{seidel} 
near the subgap resonances at
the band edge 
of the acoustical branch (at 1.5 THz) and an optical branch with the 
help of the a two-atomic chain model. The inset shows experimental results
over a wider frequency range. \label{fig8}}
\end{figure}

This might be an explanation for a resonance seen in 
Ref.\ \onlinecite{seidel} at $3.2$ mV ($\hat{=} 1.54 {\rm THz}$) 
in the $I$-$V$-characteristic of Tl$_2$Ba$_2$Ca$_2$Cu$_3$O$_{10}$, because
 the same frequency is expected by lattice dynamical calculations
\cite{kulkarnitl2} for the upper edge of the acoustical band,  
and there are no optical phonon bands in this low frequency range.
Fig.\ \ref{fig8} shows a fit of a $I$-$V$-curve calculated with the
two-atomic chain model (and some additional damping) to the experimental
results from Ref. \onlinecite{seidel}.

Similarily in BSCCO the upper edge of the longitudinal acoustical phonon band, 
which has been detected in inelastic neutron scattering at $5$ meV  
\cite{mook}, 
might correspond to a less pronounced phonon resonance in the intrinsic 
Josephson effect at $V_{\rm sg} =4.9$ mV ($\hat =$ 2.38 THz), which seems to
be invisible in optical experiments. Nevertheless at the present time this
interpretation is not yet fully conclusive, as the instrumental resolution of 
$0.65$ meV in the neutron scattering experiment is 
still rather large compared with the accuracy in the measurement of the 
$I$-$V$-curve.

Also the effect of the  two van-Hove singularities at the optical band edges 
 on the $I$-$V$-curve  as discussed above might have
been seen in the satellite structures at 5.17 THz (10.7 mV) and 5.6 
 (11.6 mV) in the $I$-$V$-curve of BSCCO \cite{wir2}. 
This interpretation is further
supported by the 
theoretical prediction of the bandwidth  $\sim 0.3 {\rm THz}$ and the fact that
the assignment of no other phonon mode is plausible.

\begin{minipage}[h]{0.47\textwidth}
\begin{table}
\caption{Comparison 
of the frequencies $f_{\rm sg}= \frac{h}{2e} V_{\rm dc}$
(in THz) of the most pronounced subgap resonances in Ref.\  
\protect\onlinecite{wir3}  and of 
infrared- and Raman active 
modes in Bi$_2$Sr$_2$CaCu$_2$O$_8$ and Tl$_2$Ba$_2$Ca$_2$Cu$_n$O$_{2n+4}$.    
\label{phonon_tab}}
    \begin{tabular}{ccccccc}
    \hline
    \multicolumn{7}{|c|}{  Subgap Resonances and Phonons in
      Bi$_2$Sr$_2$CaCu$_2$O$_8$ } \\ 
    \tableline   
    $f_{\rm sg}$ & 2.97 & 3.89 & 5.17 & 5.60 & Ref.\ \onlinecite{wir3} 
     & Josephsoneffect
    \\ \hline
    $f_{\rm LO}$ & 2.85 &      & 5.07 &      & Ref.\ \onlinecite{tsvetkov} &
    IR-reflecticity
    \\ \hline
    $f_{\rm LO}$  & 2.86 &      & 5.16 &      & Ref.\ \onlinecite{tajima} & 
     IR-reflecticity \\ \hline
    $f_{\rm TO}$ &      & 3.80 &      &     
    & Refs.\ \onlinecite{yurgenss,yurgenspriv}
    & Raman effect  \\ 
      \hline
     \multicolumn{7}{|c|}{  Subgap Resonances and Phonons in
       Tl$_2$Ba$_2$Ca$_2$Cu$_n$O$_{2n+4}$
      } \\ \hline 
    $f_{\rm sg}$ & 3.63 & 4.64 & & & Ref.\ \onlinecite{wir3} &
Josephsoneffect\tablenotemark[1]
     \\ \hline
     $f_{\rm LO}$ &      & 4.50 & & &  Ref.\ \onlinecite{tsvetkov} & 
     IR-reflecticity\tablenotemark[2]
     \end{tabular}
     \tablenotetext[1]{Tl$_2$Ba$_2$Ca$_2$Cu$_3$O$_{10}$}
     \tablenotetext[2]{Tl$_{2}$Ba$_{2}$Ca$_{2}$Cu$_{2}$O$_{8}$}
\end{table}
\end{minipage}

\section{Conclusions and Outlook}

In this paper we have developed a microscopic theory for the coupling between
Josephson oscillations and phonons in intrinsic Josephson systems like the
highly anisotropic cuprate superconductors. We determined the precise form of
the longitudinal dielectric function Eq.\ (\ref{ephon}) describing this 
coupling and obtained analytical results Eqs.\ (\ref{IV1},\ref{multi1}) for the 
$I$-$V$-curve for one and several resistive junctions. The principle
features and selection rules for phonon resonances in the $I$-$V$-curves are
illustrated with help of a simple lattice dynamical model.

We have shown that not only optical, but also acoustical phonons at the 
edge of the Brillouin zone couple to Josephson oscillations. This may
explain the structure observed in Ref.\ \onlinecite{seidel} in the 
$I$-$V$-curve occurring at an unusual low voltage (frequency), which is not 
found in
reflectivity experiments, testing transverse optical phonons at $\vec q = 0$,
and in lattice-dynamical calculations for infrared active phonons at  $\vec
q=0$. A weak satellite-structure observed in the $I$-$V$-curve of BSCCO
\cite{wir2} may be due to a double resonance from the two van-Hove
singularites of an optical branch. 

The analytical results are compared with 
results from a numerical integration of the coupled equations of motion for 
the
Josephson oscillations and phonons. For this purpose a simplified lattice
dynamical model has been used with one acoustical and one optical branch. 
It is found that in the   limit of large values of the McCumber parameter
the numerical results follow closely the analytical
solutions with the following exceptions: 1. Using a gradual change of the
bias-current, regions of the $I$-$V$-curve with negative differential
conductivity are skipped as is observed in current-biased experiments.
2. In the case of several resistive junctions, where several analytical
solutions are obtained, the numerical result
follows only one of the analytical solutions. The stability of the different
analytical solutions is currently investigated. It seems to be that the
solution which gives a minimum for the interaction energy between
polarisation and the electric field generated by the Josephson oscillations
at a given frequency is most stable.
The phonons thus serve to synchronise the Josephson oscillations in different  
resistive layers, which is important for the application of such systems as
high-frequency devices.

In this paper we have considered only the case of a current
distribution in the c-direction which is homogeneous along the
layers. We neglected all magnetic field effects assuming that all quantities
are uniform along the layers. For real systems this is an
artificial approximation, because the current induced magnetic field can never 
be avoided completely.
 Nevertheless, we argue that this is a valid approximation for the
intrinsic Josephson systems forming a mesa structure of about fifty layers
with a width of several $\mu$ and a thickness which is much smaller. Such 
a junction is still short with respect to the length $\lambda_J= c/\omega_p$ 
describing the variation of the phase-difference along the layers (in order to
avoid completely the spontaneous formation of vortices, which have a smaller
diameter, the mesa width should not exceed 1-2 $\mu$).   Therefore
the treatment of the superconducting layers as metal sheets with a uniform
charge distribution, the creation of uniform polarisation fields and the 
neglect of $q_\parallel$ in the calculation of the longitudinal dielectric
function is justified for the systems investigated. 
This will be different in the case of longer
junctions and strong external magnetic field \cite{wir6}.
In particular the flow of vortices and their interaction with phonons has to be
investigated in this case \cite{wir5}.

\section*{Acknowledgement}

The authors would like to thank A. Yurgens and A. Tsvetkov for discussions
on their experimental results and for providing unpublished data, A. Mayer 
and D. Strauch for fruitful discussions on lattice dynamical aspects, 
and P. M\"uller, R. Kleiner, L. Bulaevskii and A. Bishop for the continuous 
support of our work on  intrinsic Josephson systems. Financial support by the
Deutsche Forschungsgemeinschaft, the Bayerische Forschungsstiftung within the
research project FORSUPRA,  the Studienstiftung des Deutschen Volkes (C.H.) 
and the US Department of Energy under contract W-7405-ENG-36 (C.H.) is 
gratefully acknowledged.

\begin{appendix}

\section{$I$-$V$-curves for several resistive junctions}

Quite generally the RSJ-equation for the n-th junction can be written as        
\begin{eqnarray}\label{RSJ2}
j &=&j_c \sin\gamma_n(t) + \sigma \frac{\hbar}{2ed} \dot \gamma_n(t) \\
\nonumber
&+&\frac{\epsilon_0\hbar}{2ed} \sum_{n'} \int_{-\infty}^\infty
\epsilon^L_{\rm ph}(n - n',t-t') \ddot \gamma_{n'}(t') dt'. 
\end{eqnarray}
where $\epsilon^L_{\rm ph}(n-n',t)$ is the dielectric function 
Eq.\ (\ref{ephon}) in real
space and time domain. 

Let us denote the index of a general junction by $n \in \{1 \dots N \}$ 
and the subset
of indices of resistive junctions by $I$. In the following we assume that
oscillations in all junctions are governed by one frequency $\omega$  and the
phases can be written as 
\begin{equation}
\gamma_n(t) = \cases{\theta_j + \omega t + \delta \gamma_j(t)
& $n=j\in I$ \cr
\theta_n + \delta \gamma_n(t) & else \cr}
\end{equation}
with  
\begin{equation}
\delta \gamma_n(t) = \delta \gamma_ne^{-i\omega t} + c.c.
\end{equation}
Expanding with respect to $\delta \gamma_n$ and keeping only the
lowest harmonic one finds 
\begin{equation}
\sin\gamma_n(t) \simeq \cases{
\sin(\theta_j + \omega t) + \Re (\delta \gamma_j e^{i\theta_j}) & $n=j\in I$
\cr \sin\theta_n + \cos \theta_n \delta \gamma_n(t) & else. \cr}
\end{equation}
Splitting the RSJ-equation (\ref{RSJ2}) into a dc-part and an oscillating part
one obtains for the dc-part:
\begin{equation}
j= \cases{ \sigma E_{dc} + j_c \Re ( \delta \gamma_j e^{i\theta_j}) & for
$n=j\in I$ \cr
j_c \sin\theta_n & else, \cr } \label{j}
\end{equation}
while the oscillating part can be written as
($\bar\omega_p^2=\omega_p^2\sqrt{1-(j/j_c)^2})$
\begin{eqnarray}
\bar \omega_p^2 \delta \gamma_n(t) &+& \frac{\sigma}{\epsilon_0}
\delta \dot \gamma_n(t)\\ \nonumber
&+& \sum_{n'} \int_{-\infty}^\infty \epsilon(n-n',t-t')
 \delta \ddot \gamma_{n'}(t')dt' = f_n(t) 
\end{eqnarray}
with
\begin{equation}
f_n(t)=\cases{\bar \omega_p^2 \delta \gamma_j(t) - \omega_p^2 \sin(\theta_j +
\omega t) & $n=j \in I $\cr
0& else.\cr }
\end{equation}
Writing 
\begin{equation}
f_j(t) = f_je^{-i\omega t} + c.c.
\end{equation}
the amplitude of the driving term for a resistive junction is given by  
\begin{equation}
f_j = \bar\omega^2_p \delta \gamma_j - i \frac{\omega^2_p}{2}
e^{-i\theta_j}.
\end{equation}

Introducing the spatial Fourier transform 
\begin{equation}
\gamma(q_z) = \frac{1}{N_z} \sum_{n=1}^{N} \delta \gamma_n e^{-iq_z z_n}
\end{equation}
we obtain
\begin{eqnarray}
\gamma(q_z) &=& G(q_z,\omega) \sum_{j\in I} f_j e^{-iq_zz_j} \\ \nonumber
&=& G(q_z,\omega ) \sum_{j\in I} \Bigl[ \bar \omega^2_p \delta \gamma_j - i
\frac{\omega^2_p}{2} e^{-i\theta_j}\Bigr] e^{-iq_zz_j}
\end{eqnarray} 
with the Green's function 
\begin{equation}
G^{-1}(q_z,\omega) = \bar\omega_p^2 - i\omega \frac{\sigma}{\epsilon_0} -
\omega^2\epsilon^L_{\rm ph}(q_z,\omega)
\end{equation} 
of the homogeneous equation. 

From this an equation for $\delta \gamma_i$ ($i \in I$) 
in the resistive junctions is obtained:
\begin{equation}
\sum_{j\in I}(G^{-1}(i,j,\omega ) - \bar \omega^2_p\delta_{i,j} )\delta
\gamma_j = - i \frac{\omega^2_p}{2} e^{-i\theta_i}
\end{equation}
with
\begin{equation}
G(i,j,\omega)= \frac{1}{N_z}\sum_q G(q,\omega) e^{iq(z_i-z_j)}.
\end{equation}
Using 
\begin{equation}
\bar\epsilon_{\rm ph}(i,k,\omega) = -\frac{1}{\omega^2} G^{-1}(i,k,\omega)
+  (\frac{\bar\omega^2_p}{\omega^2}
-\frac{i\sigma}{\epsilon_0 \omega} ) \delta_{i,k} 
\end{equation}
we get
\begin{equation}
\delta \gamma_i = i\frac{\omega^2_p}{2 \omega^2}\sum_{k \in I}
\Bigl[\bar \epsilon_{\rm ph}(i,k,\omega) +
\frac{i\sigma}{\epsilon_0\omega}\delta_{i,k}\Bigr]^{-1} 
e^{-i\theta_k},  
\end{equation}
which gives us the dc-component Eq.\ (\ref{j}) of the current density           
\begin{eqnarray}
j(V) &=& \sigma E_{dc} \\ \nonumber
&-&j_c \frac{\omega^2_p}{2
\omega^2} \Im \sum_{k \in I} e^{i\theta_i}\Bigl[\bar 
\epsilon_{\rm ph}(i,k,\omega)
+ \frac{i\sigma}{\epsilon_0\omega}\delta_{i,k}\Bigr]^{-1} 
e^{-i\theta_k}.  
\end{eqnarray}

Finally we want to note that the dielectric function $\bar
\epsilon_{\rm ph}(i,k,\omega)$ can also be used to write the RSJ equations in a
form where only the phases $\gamma_i(t)$, of the resistive junctions $i \in I$ 
enter: 
\begin{eqnarray} 
j =j_c \sin\gamma_i(t) &+& \sigma \frac{\hbar}{2ed} \dot \gamma_i(t) \\
\nonumber
&+&\frac{\epsilon_0\hbar}{2ed} \sum_{j\in I} \int_{-\infty}^\infty \bar
\epsilon_{\rm ph}(i,j,t-t') \ddot \gamma_j(t') dt'.  
\end{eqnarray}

\end{appendix}

\end{document}